# A HISTORICAL NOTE ABOUT HOW THE PROPERTY WAS DISCOVERED THAT HYDROGENATED SUBSTANCES INCREASE THE RADIOACTIVITY INDUCED BY NEUTRONS

*Alberto De Gregorio*
*Doctorate of research in Physics*
Università di Roma «La Sapienza»[*]

At the *Domus Galilæana* in Pisa, many original documents and records are kept, which belong to the scientific activity carried out by Enrico Fermi until 1938. Such documents range from Fermi's very first works – including the manuscript of the thesis for his degree – to some notes written a few months before Fermi's departure for the United States. Such papers are of the highest importance in analysing the activity which Fermi carried out while in Italy.

## 1. – Experiments on neutrons and the discovery of the property of hydrogenated substances

The Roman group headed by Fermi, which included Edoardo Amaldi, Oscar D'Agostino, Bruno Pontecorvo, Franco Rasetti and Emilio Segrè, worked very hard on artificial radioactivity induced by neutrons from March to October 1934 and even later. It is not surprising, then, what a vast number of original papers the *Domus Galilæana* offers as a documentation of that fundamental scientific work.

In carrying out my work for the Doctorate of research in Physics – headed by professor Fabio Sebastiani – I compared those documentary sources with the supported evidences, the personal recollections, concerning the pioneering researches on radioactivity induced by neutrons. Such a comparison was restricted, actually, only to the discovery of the property of hydrogenated substances, i.e., was restricted to only a part of the whole of the researches carried out in Rome in 1934. Still, even at such a preliminary analysis, very interesting conclusions can be drawn from the documents kept at the *Domus Galilæana*.

---

[*] alberto.degregorio@roma1.infn.it

*1.1. – Towards radioactivity induced by neutrons*

As early as 1919, Ernest Rutherford discovers that nitrogen nuclei emit protons, if they are irradiated by α-particles. Still in the early Thirties, disintegrations produced by α-particles are widely investigated, among others, by James Chadwick. On February 17th 1932, the latter announces the *Possible Existence of a Neutron*. Just the day before, Norman Feather observed some traces in a cloud chamber, produced by the penetrating radiation from beryllium. The following week, those traces are correctly interpreted as due to nitrogen nuclei disintegrated by neutrons from beryllium. Rutherford announces Feather's discovery before the Royal Society on March 18th; in May, that discovery is published in the *Proceedings*.[1]

At that time, only weak sources of neutrons were available – only one neutron is emitted when beryllium is irradiated with about 100,000 α-particles. Nevertheless, Feather himself observes that of 130 cases of interaction between a neutron and a nitrogen nucleus, about 30 result in disintegration; in the case of α-particles, instead, elastic collisions outnumber inelastic collisions by a factor of the order of 1000:1. That Feather ascribes to «the different extent of the external fields of the two particles».

In January 1934, the first signs of the discovery having been discussed since the Solvay Conference the previous October, Frédéric Joliot and Irène Curie show that α-particles can induce radioactivity in some light elements – the first they activated were boron, magnesium and aluminium. The Joliot-Curie themselves conclude that the acrtivation could possibly occur also by means of some different kinds of particles.

Fermi, then, guesses that neutrons can not only produce disintegrations – like those Feather observed – but could also produce artificially radioactive elements. Neutrons, moreover, due to their electrical neutrality, could also activate heavy nuclei. On March 25th 1934, Fermi announces he has observed artificial radioactivity induced by neutrons.[2] Since then, the Roman group headed by Fermi systematically investigate radioactivity induced by neutrons, getting as far as investigating uranium. The very experiments with uranium let Fermi believe – with some hesitations, actually – he has discovered two new elements with atomic numbers 93 and 94. Concerning that, it might be interesting to report that the documents kept in Pisa show that, since May 1934, it became a habit in Fermi's laboratories to use the symbol «Ao» to signify the Ausonium, the hypothetical new element of atomic number 93; for Hesperium, of atomic number 94, the symbol «Hs» was used. That seems to contradict Segrè, who recalls that, after the Director of the Institute Orso Mario Corbino had delivered a speech before the *Accademia dei Lincei* on June 4th, Fermi «strongly resisted the temptation to give new names to the so-called transuranic elements».[3]

*1.2. – Personal recollections of a new discovery*

At first, the activity induced is classified as «weak», «medium» and «strong». As Amaldi and Segrè recall,[4] for the sake of a finer investigation the activity with a two minutes half-life induced in silver is soon chosen as a test activity. At this point, Fermi and his group notice that the environment seems to affect the activation, and they decide to elucidate what really occurs. To this end, they put some lead near the source of the neutrons; moreover, they decide to interpose the lead as a shield, between the source and the target to be activated.

In 1984, Amaldi so recalls what happened fifty years before: «On the morning of October 22$^{nd}$ most of us were busy doing examinations and Fermi decided to proceed in making the measurements». Still, according to Amaldi, Fermi suddenly replaced the lead, he had not used yet, with some odd piece of paraffin: as an unexpected result, the activity increased by an appreciable amount. Fermi promptly came to an explanation for what had happened: neutrons slowed down by hydrogenated substances increase their efficiency. «The same afternoon the experiment was repeated in the pool of the fountain in the garden of the Institute. […] The evening of October 22$^{nd}$ all the group came to my house and a letter announcing our results was written to *La Ricerca Scientifica*», Amaldi writes.[5]

There is no doubt at all, according to Amaldi's account, that everything happened on Monday, October 22$^{nd}$ 1934: the discovery of the effects due to the paraffin, the interpretation of those experiments, the experiments with water, as well as the very writing the article for *La Ricerca Scientifica*.

Before Amaldi, in 1970 Segrè wrote: «[…] a lead wedge was prepared for insertion between the neutron source and the detector […] but Fermi suddenly decided to try filters of light elements first. […] It is very hard for me, after so many years, to remember exactly what happened, but there is no doubt that paraffin was used first on the morning of October 22».[6] *Paraffin was used first on the morning of October 22…*

In 1955, i.e. twenty-one years after those events, Persico recalled that on the morning of October 22$^{nd}$ 1934, Fermi's group realised that the radioactivity induced in a target of silver increased a lot, if some paraffin was placed nearby.[7] Pontecorvo as well – who joined the group in September 1934, shortly after his degree – in 1972 says: «[…] On the morning of October 22$^{nd}$ 1934, Fermi made up his mind to measure the radioactivity of a cylinder of silver, by passing the neutrons from the source through a ledge of paraffin he had rapidly prepared, instead of passing them through a ledge of lead».[8]

A laconic account is given by Rasetti in 1968: «*Nell'autunno [del 1934] si aggiunse al nostro gruppo Pontecorvo, e tosto scoprimmo gli effetti sorprendenti che certe sostanze, l'acqua e la paraffina, producevano nell'intensificare la radioattività indotta quando si trovassero nelle vicinanze della sorgente di neutroni e dell'elemento bombardato. Non passò* *un giorno che Fermi aveva già trovato la spiegazione di questi effetti paradossali nel rallentamento che i neutroni subiscono urtando più volte contro i nuclei di idrogeno contenuti nell'acqua o simili sostanze*».[9]

Of all the accounts quoted here, the first one in a chronological order is provided by Fermi's wife, Laura Capon. In 1954, she published a biography of Fermi, in which she recalls: «A plate of lead made the activity increase slightly. Lead is a heavy substance. "Let's try a light one next," Fermi said, "for instance, paraffin." The experiment with paraffin was performed on the morning of October 22».[10]

Each account differs from the others, and has its own shade of meaning. Nevertheless, apart from Rasetti's account, in which no date is mentioned at all, all the others come to an agreement, in saying that Fermi used paraffin on October 22nd 1934.

He, whose account does not echo the others, is D'Agostino, the chemist of the group. In a controversial reconstruction of those events, D'Agostino ascribes Fermi's discovery of the effect due to hydrogenated substances, to the presence of a pail with water, rather than to the presence of some odd piece of paraffin: the charwoman used to leave the pail beneath the bench, upon which Pontecorvo generally carried out his experiments. D'Agostino recalls: «*Sì, era vero: l'attività dell'argento radioattivo variava a seconda che nel secchio c'era o non c'era l'acqua. Palesemente eccitato Fermi suggerì ancora: "Proviamo a immergere l'argento in una grande quantità d'acqua!". Era lì a portata di mano la fontana del giardino [...], in un cortile interno dell'Istituto. Erano le ore 15 del 22 ottobre. [...] L'acqua moltiplicava la radioattività dell'argento di un fattore assai grande*».[11]

Even though D'Agostino provides an account which totally disagrees with the others, he still asserts that it was October 22nd. Yet, further on in his report a detail appears, which is inconsistent with such a date: according to D'Agostino, in fact, he himself suggested that the discovery of the property of hydrogenated substances should be patented. Some in the group were sceptical about that suggestion. Nevertheless, Fermi settled to refer any decision to the Director of the Institute, senator Corbino. «*Ricordo anche che queste nostre discussioni avvennero di sabato. Il senatore Corbino era fuori Roma e sarebbe tornato soltanto due giorni dopo. Così ogni decisione fu rimandata al lunedì*».[12] The *brevetto* [patent] number 324458 was asked through a letter dated October 26th 1934: it was Friday.[13] Then, according to the episode which D'Agostino here reports, the discovery of the effect of hydrogenated substances necessarily occurred no later than Saturday, October 20th.

Despite the more or less marked discrepancies revealed in all the accounts reported above, the agreement about the date is really complete. To the point that it seems even suspicious: we saw that, according to Segrè, it was «very hard […], after so many years, to remember exactly what happened». Is it really possible that, in spite of such a difficulty, all of them – except Rasetti, if any, and in spite of the inconsistencies which D'Agostino's report holds – could exactly remember the date in which Fermi did discover the properties of hydrogenated substances? Moreover, Fermi himself failed to mention October 22[nd] in his only known recalling of that discovery, which he related to Subrahmanyan Chandrasekhar between Autumn 1952 and Spring 1953:[14] a bit singular is the very fact that, in spite of this, all the other accounts – again, except Rasetti's one – do state precisely that date.

The matter gains a particular importance in the light of the documents held at the *Domus Galilæana*.

*1.3. –Records at the archives of the* Domus Galilæana

Amaldi clearly states that the results of the first experiments with paraffin were recorded on notebook *B.1*, nowadays kept in Pisa.[15] In the very first pages, some measurements are recorded, which were obtained in presence of the «*castelletto*» [«little castle»], a small squared fencing made of small bricks of lead: it was used to investigate the effects of the environment on the activation of silver. Starting from page 8, the effects on that activation, due to paraffin, are investigated. Anyway the date of October 20[th] '34 – instead of October 22[nd] – is clearly written at the top of the page (**fig. 1**). One could perhaps think of an error. Nevertheless, some more records confirm that the experiments took place on October 20[th]. Furthermore, they also add some more details of interest.

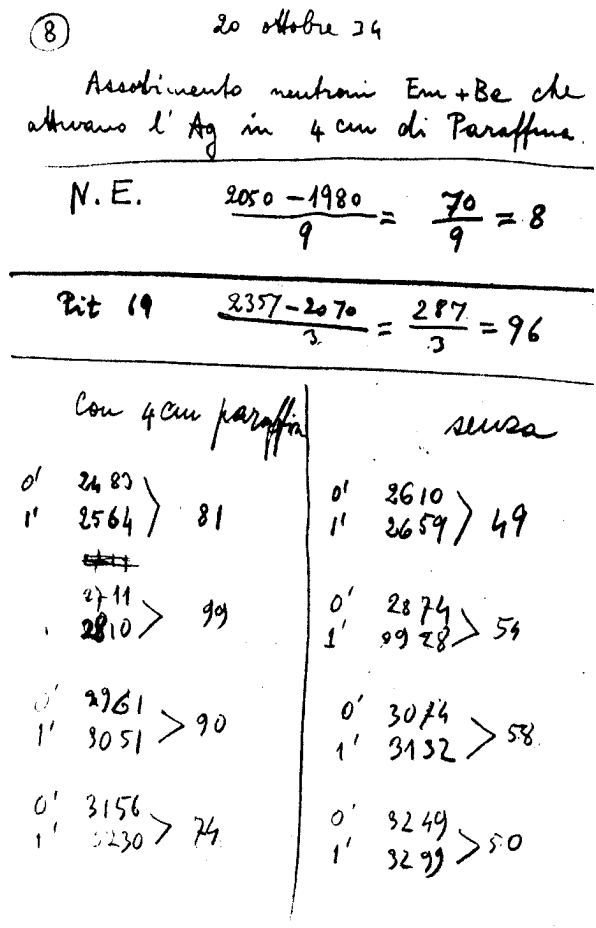

Figure 1. – Page 8 of notebook *B.1*, dated October 20[th] 1934, recording experiments with paraffin.

In *Report R.6*, many graphs and about ninety «verbali» – i.e. imprinted forms to record the results of the experiments – are kept. They record the results of the experiments worked out since October 1934. One of these *verbali*, dated October 21$^{st}$ 1934, reports that since 20 (8 p.m.) of Saturday, October 20$^{th}$, until 9.30 a.m. of Sunday, October 21$^{st}$, caesium is irradiated while soaked under some water; the number is also reported of the radon-beryllium source of neutrons used: source number 16. The radioactivity induced in caesium is measured, and then reported. Another «verbale» reports that since 20 of October 20$^{th}$, till 9.30 of the following morning, rubidium nitrate ($RbNO_3$) is irradiated; the number of the source of neutron is not specified, but the agreement on the hours seems to suggest that the source is the same used for caesium: source number 16.

So, caesium was irradiated while soaked under water as from the evening of October 20$^{th}$: that was the same day when, according to notebook *B.1*, the experiment with paraffin was carried out.

An approximate check of the dates is possible, if we refer to the numbers of the sources of neutrons used. In *Report R.9*, in fact, the numbers and the activities are reported of the sources that, day by day, are used in the experiments with neutrons (**fig. 2**). Those activities decay following an exponential law, with half-life of about four days. Unfortunately, the only data missing are precisely those referring to the activity of source number 16 – the reason perhaps being that, in those days, Fermi's group was too deeply involved in the real experiments to be concerned with the activity of the source. Such a difficulty is not serious, indeed, because the sources are progressively numbered according to a chronological ordering: by only considering the dates when the other sources were prepared, we can be quite sure that source number 16 could be used by Fermi and his group in about the middle of October 1934. In other words, though it is not possible to reconstruct day by day the activity of the source used to irradiate caesium, it is still possible to reconstruct from *Report R.9* that source number 16 could provide an effective flux of neutrons just spanning October 20$^{th}$. This confirms the trustworthiness of the date reported on the record about caesium, and it excludes, for example, that the experiment with caesium was carried out on November 20$^{th}$: at that date, only the activity of the sources number 19 (60 mCi), 20 (220 mCi) and 21 (440 mCi) were still enough for experimental purposes.

|      | 14  | 15  | 16  | 17  | 18  |
|------|-----|-----|-----|-----|-----|
| 1-X  | 285 |     |     |     |     |
| 2    | 240 |     |     |     |     |
| 3    | 200 |     |     |     |     |
| 4    | 168 |     |     |     |     |
| 5    | 140 |     |     |     |     |
| 6    | 117 |     |     |     |     |
| 7    | 97  |     |     |     |     |
| 8    | 82  | 215 |     |     |     |
| 9    | 68  | 180 |     |     |     |
| 10   |     | 150 |     |     |     |
| 11   |     | 125 |     |     |     |
| 12   |     | 105 |     |     |     |
| 13   |     | 88  |     |     |     |
| 14   |     |     |     |     |     |
| 15   |     |     |     |     |     |
| 16   |     |     |     |     |     |
| 17   |     |     |     |     |     |
| 18   |     |     |     |     |     |
| 19   |     |     |     |     |     |
| 20   |     |     |     |     |     |
| 21   |     |     |     |     |     |
| 22   |     |     |     | 430 |     |
| 23   |     |     |     | 350 |     |
| 24   |     |     |     | 290 |     |
| 25   |     |     |     | 240 |     |
| 26   |     |     |     | 200 |     |
| 27   |     |     |     | 170 |     |
| 28   |     |     |     | 140 |     |
| 29   |     |     |     | 120 | 365 |
| 30   |     |     |     |     | 305 |
| 31   |     |     |     |     | 255 |

|      | 18  | 19  | 20  | 21  |
|------|-----|-----|-----|-----|
| 1-XI | 215 |     |     |     |
| 2    | 180 |     |     |     |
| 3    | 150 |     |     |     |
| 4    | 125 |     |     |     |
| 5    | 105 |     |     |     |
| 6    | 88  | 750 |     |     |
| 7    | 73  | 630 |     |     |
| 8    | 61  | 525 |     |     |
| 9    | 51  | 440 |     |     |
| 10   | 43  | 365 |     |     |
| 11   | 36  | 305 |     |     |
| 12   | 30  | 255 |     |     |
| 13   | 25  | 215 | 745 |     |
| 14   |     | 175 | 625 |     |
| 15   |     | 146 | 525 |     |
| 16   |     | 123 | 440 |     |
| 17   |     | 103 | 372 |     |
| 18   |     | 86  | 312 | 624 |
| 19   |     | 72  | 260 | 520 |
| 20   |     | 60  | 220 | 440 |
| 21   |     |     | 185 | 370 |
| 22   |     |     | 155 | 310 |
| 23   |     |     | 131 | 262 |
| 24   |     |     | 110 | 220 |
| 25   |     |     | 92  | 184 |
| 26   |     |     | 78  |     |
| 27   |     |     | 65  |     |

Figure 2. – From *Report R.9*: *facsimile* of the record of the sources of neutrons, concerning the months of October and November 1934. In each column, the number of the source and the daily activity of the latter are recorded.

Going on with our analysis of the records contained in *Report R.6*, we find aluminium «*irradiato in H₂O a 10 cm per 2'*» on October 21st. Since 20 of Sunday, October 21st, some more substances are irradiated while soaked under water: sodium carbonate ($Na_2CO_3$), until 9.45 a.m. of Monday, October 22nd; lithium hydroxide (LiOH), until 10 a.m.; platinum until 10.30 a.m. **(fig. 3)**; ruthenium and strontium until 12 a.m. In each of these cases, it is clearly stated that 1) irradiation took place in water and 2) the source of neutrons used was source number 16. The latter point makes us conclude that, being the source one and the same, possibly all of those samples were soaked under water the one next to the other, while irradiated. On Monday,

October 22$^{nd}$, by means of source number 16, the following substances are also irradiated: «*in H$_2$O Pb metallico*» – between 13 and 15 –, and metallic antimony «*in provetta in H$_2$O*» – in a test-tube, between 13 and 16.20.

Figure 3. – From *Report R.6*: record about platinum, which was irradiated, like other substances, while soaked under water during the night between October 21$^{st}$ and October 22$^{nd}$.

On October 20$^{th}$, calcium fluoride (CaF$_2$) turns out to have been irradiated, apparently without water – in any case, calcium fluoride was largely investigated in the preceding months. Ammonium iodide (NH$_4$I) as well turns out to be irradiated on October 20$^{th}$, between 13 and 14, in a test-tube containing some water. Anyway, in these last two cases, the real date is not so sure, since the number of the source is not stated. On the other hand, ammonium nitrate (NH$_4$NO$_3$) turns out to have been irradiated on October 15$^{th}$, while potassium carbonate (K$_2$CO$_3$) on October 16$^{th}$. Those last two dates, however, are clearly wrong, since the sources used were number 19 and 20: they would be active only since the middle of the following November.

## 2. – Conclusions

We can thus come to the conclusion that the experiments with paraffin were carried out on October 20$^{th}$ 1934, as reported on notebook *B.1*, and not on October 22$^{nd}$, as has been believed so far. Moreover, exactly the same procedure was worked out during the following two nights: since the evening of that October 20$^{th}$, some specimens were soaked under water and irradiated during the following night, and the same happened on the night between 21st and 22nd. The procedure has an odd analogy with D'Agostino's account, according to which the pail with water, that the charwoman used to leave beneath Pontecorvo's bench, was «*messo alla sera e ritirato poi al mattino*» [«placed in the evening and then taken away in the morning»].

The documents kept at the *Domus Galilæana*, then, contradict what Fermi's collaborators and wife reported.

One probable check for our conclusion is annotated on one record dated November 20th 1934, kept in *Report R.1*. There are recorded the results obtained with sodium carbonate, irradiated alternately in air and in paraffin – using source number 20. There is reported that «*Le misure giusti* [sic]*, senza sbagli geometrici, sono quelle del 21.10.34*» [«The correct measurements, without geometrical mistakes, are those of 21.10.34»] **(fig. 4)**.

Figure 4. – From *Report R.1*: results of experiments carried out – with some paraffin – in November, in order to check some earlier results which dated back to October 21st.

One could perhaps object, that as many as six accounts agree on the date of October 22$^{nd}$. Nevertheless, if it is certainly true that a whole group of persons reconstructed those events, it is likewise true that the experiments were carried out and the results were recorded at the presence of that same group of persons. There is no reason to believe that all of them were wrong, when writing on the records the date of October 20$^{th}$ and then the date of October 21$^{st}$ – even if some inaccuracy really occurs elsewhere –, while they were right when, some decades later, they reported the date of October 22$^{nd}$. It is really possible, rather, that the mistake first occurred perhaps in the biography written by Laura Fermi in 1954, and then 'propagated' through Persico's account, to those of all the others: such a circumstance could also explain the inconsistencies in D'Agostino's report.

The shift of two days in dating those events is not so much important in itself. Its major importance is, instead, that such a shift shows that the accounts that have been supported so far, and regarding the experiments carried out at via Panisperna by Fermi and his group, must be examined closely: such accounts must be carefully analysed and compared with the archive records, even there, where they show a complete agreement among themselves. Any historical study to come, concerning the experiments carried out by Fermi and his group on neutrons, cannot neglect such a check.

*Addendum*

It is worth noting that, according to the pieces of information which emerge from the catalogue of Fermi's documents at the *Domus Galilæana*[16] – and which seem to have apparently been confirmed by now, by the preliminary analysis I carried out –, we can conclude that any sort of Fermi's documents break up on March 26$^{th}$ 1938. The real interest in noting such a detail is that March 26$^{th}$ 1938 is precisely the date when Ettore Majorana disappeared. The question is worth being examined closely.

---

[1] N. Feather, *Collisions of Neutrons with Nitrogen Nuclei*, «Proc. Roy. Soc.», A CXXXVI (1932), pp. 709-727.

[2] E. Fermi, *Radioattività indotta da bombardamento di neutroni*, «La Ricerca Scientifica», 5, 1 (1934), p. 283; in E. Fermi, *Note e memorie (Collected Papers)*, ed. by Edoardo Amaldi *et al.*, Roma, Accademia Nazionale dei Lincei - Chicago, the University of Chicago Press, 1962-1965, 2 vols.; vol. I, pp. 645-646. For a wider account of the researches worked out with neutrons in Rome, see for example: L. Carbonari and F. Sebastiani, *La scoperta dell'azione di sostanze idrogenate sulla radioattività provocata da neutroni: nota in margine al ritrovamento di un cimelio fermiano*, to be published on «Physis»; F. Cordella *et al.*, *Enrico Fermi. Gli anni italiani*, Roma, Editori Riuniti, 2001.

[3] E. Segrè, *Enrico Fermi. Physicist*, Chicago, The University of Chicago Press, 1970, p. 77.

[4] E. Amaldi, *From the Discovery of the Neutron to the Discovery of Nuclear Fission*, «Physics Reports», 111, 1-4, pp. 151-152; Segrè's account is reported in his introduction to Fermi's articles about artificial radioactivity, in E. Fermi, *Note e memorie (Collected Papers)*, quoted, vol. I, pp. 641-642.

[5] E. Amaldi, *From the Discovery of the Neutron*, quoted, pp. 152-154.

[6] E. Segrè, *Enrico Fermi. Physicist*, quoted, pp. 79-80.

[7] E. Persico, *Souvenir de Enrico Fermi*, «Scientia», XC, 1955, pp. 319-320.

[8] B.M. Pontecorvo, V.N. Pokrovskij, *Enrico Fermi v vospominanijakh uchenikov i druzej*, Moskva, Nauka, 1972 [Italian transl., B. Pontecorvo, *Enrico Fermi*, Pordenone, Edizioni Studio Tesi, 1972, pp. 81-82. The Italian translation, from wich the English version here reported is drawn, reads: «*La mattina del 22 ottobre 1934 Fermi decise di misurare la radioattività del cilindro d'argento 'facendo passare' i neutroni dalla sorgente attraverso un cuneo non di piombo, ma di paraffina che lui stesso aveva in fretta approntato*»].

[9] *In Autumn [1934], Pontecorvo joined our group. We soon discovered the surprising effects that some substances like water or paraffin have, when they are placed near the source of neutrons and near the bombarded element, of increasing the radioactivity they induce. Within not even one day, Fermi explained those paradoxical effects, and ascribed them to the slowing down suffered by neutrons when colliding repeatedly against hydrogen nuclei contained in water or similar substances.* Author's translation into English, from the original: F. Rasetti, *Enrico Fermi e la fisica italiana*, in *Celebrazioni Lincee in ricordo di Enrico Fermi*, Roma, Accademia Nazionale dei Lincei, 1968, pp. 13-14.

[10] L. Fermi, *Atoms in the family. My life with Enrico Fermi*, Chicago, The University of Chicago Press, 1954, p. 98.

[11] *Yes, it was really true: the activity of the radioactive silver changed, depending on the fact that some water was, or instead was not inside the pail. Visibly excited, Fermi moreover suggested: «Let's try to soak the silver in a large quantity of water!». Within reach was the fountain in the garden of the Institute. It was 15 o' clock of October 22$^{nd}$. […] Water increased the activity induced in silver by a very large amount.* Author's translation, from: O. D'Agostino, *L'era atomica incominciò a Roma nel 1934*, 2$^{nd}$ episode, «Candido», year XVI nr. 24, June 15$^{th}$ 1958, pp. 24-25.

[12] *I can remember, as well, that those discussions of ours took place on Saturday. Senator Corbino was not in Rome, and would be back only within two days. For this reason, any sort of decision was postponed to the next Monday*. Author's translation, from: O. D'Agostino, *L'era atomica incominciò a Roma nel 1934*, 3$^{rd}$ episode, «Candido», year XVI nr. 25, June 22$^{nd}$ 1958, p. 21.

[13] A copy of this letter can be consulted on the Web site of the Museum of the Department of Physics, University of Rome *La Sapienza*: www.phys.uniroma1.it/DOCS/MUSEO/home.htm.

[14] Fermi's account was quoted by Chandrasekhar in E. Fermi, *Note e memorie (Collected Papers)*, quoted, vol. II, pp. 926-927.

[15] The classification of the records kept in Pisa, to which I refer, is the one reported in M. Leone, N. Robotti, C.A. Segnini, *Fermi Archives at the* Domus Galilæana *in Pisa*, «Physis», XXXVII, 2 (2000), pp.501-533. It is worth noting that the documentation kept at the *Domus Galilæana*, and concerning the experiments carried out on uranium by Fermi and his group, appears to be incomplete: *Report R.2* contains a few hundreds of numbered records, but such a numeration is repeatedly interrupted, in a way which clearly shows that some more hundreds of records are missing.

[16] Ibid.